\newcommand{\Msun}{~M_\odot}

\newcommand{\gcm}{\rm ~g~cm^{-3}}
\newcommand{\kms}{\rm ~km~s^{-1}}

\documentclass[12pt,preprint]{aastex}

\begin{document}


\title {\bf  BULLETS IN A CORE COLLAPSE SUPERNOVA REMNANT:
THE VELA  REMNANT} 

\author{Chih-Yueh Wang\altaffilmark{1} and Roger A. Chevalier}
\affil{Department of Astronomy, University of Virginia}
\affil{P.O. Box 3818, Charlottesville, VA 22903-0818}
\email{cw5b@virginia.edu, rac5x@virginia.edu}

\altaffiltext{1}{Present address: 
Institute of Astronomy, National Central University,
Chungli, Taiwan 320}
\vspace{1.0cm}

\begin{abstract}

We use two-dimensional hydrodynamical simulations to investigate 
the properties of dense ejecta clumps (bullets) in a core
collapse supernova remnant, motivated by the observation of 
protrusions probably caused by clumps in
the Vela supernova remnant.
The ejecta, with an inner flat and an outer steep power law density
distribution, were assumed to freely expand into an ambient medium 
with a constant density,
$\sim 0.1$ H atoms cm$^{-3}$ for the case of Vela. 
At an age of $10^4$ yr,
the reverse shock front is expected to have moved back to the
center of the remnant.
Ejecta clumps  with an initial density 
contrast $\chi \sim 100$ relative to their
surroundings  are found to
be rapidly  fragmented and decelerated.   
In order to cause a pronounced protrusion on the blast wave,  as observed
in the Vela remnant,
$\chi \sim 1000$ may be required.
In this case, the clump should be near the inflection point in
the ejecta density profile, at an 
ejecta velocity $\sim 3000 \kms $.
These results apply to moderately large clumps; smaller clumps would
require an even larger density contrast.
Clumps can create ring structure in the shell of the Vela remnant and
we investigate the possibility that RX J0852--4622, an apparent 
supernova remnant
superposed on Vela, is actually part of the Vela shell.
Radio observations support this picture, but the possible presence
of a compact object argues against it.
The Ni bubble effect or compression in a pulsar wind nebula are
 possible mechanisms to produce the clumping.

\end{abstract}

\keywords{hydrodynamics ---  
           supernova remnants 
          --- supernovae: general --- ISM: individual (Vela supernova remnant)}


\section {INTRODUCTION}

There is widespread evidence that the ejecta of core collapse supernovae
are clumpy.
The oxygen line profiles in the nearby Type II supernovae SN 1987A
(Stathakis et al. 1991) and SN 1993J (Spyromilio 1994;
Matheson et al. 2000) showed 
evidence for structure, implying
that the gas is clumped.
The velocity range for the emission extends to 
$1,500\kms$ in SN 1987A and $-4,000\kms$
in SN 1993J.
Similar evidence for clumping has  been found in the Type Ib
supernova SN 1985F (Filippenko \& Sargent 1989).
Among young supernova remnants, Cas A is the prototype of the
oxygen-rich SNRs, which show evidence for freely expanding, oxygen-rich
ejecta in clumps
(Chevalier \& Kirshner 1979).
Puppis A, with an age $\sim 3,700$ yr, is a more elderly example of such
a remnant (Winkler et al. 1988).
The finding of a neutron star in the remnant demonstrates that it
is a core collapse supernova (Pavlov, Zavlin, \& Tr\"umper 1999).

Clumping of the ejecta is also found in the Crab Nebula, in which
the ejecta have been swept up and compressed by a pulsar wind bubble.
When a remnant with clumpy ejecta runs into the surrounding medium,
the clumps can be expected to move ahead of the blast wave created by
the smooth ejecta.
An apparent case of such evolution involves X-ray and radio
observations of the Vela supernova remnant (SNR).
The finding of protrusions around the periphery of the remnant is
best explained by ejecta clumps (Aschenbach, Egger, \& Tr\"umper 1995;
Strom et al. 1995).
X-ray spectra of one of the protrusions shows evidence for an enhanced
Si abundance, confirming the ejecta clump picture (Tsunemi, Miyata, 
\& Aschenbach 1999).
Kundt (1988) had discussed the possibility that supernovae are
explosions of `shrapnel,' giving rise to a complex outer boundary.

Another development
 is the finding of an apparent young remnant superposed on the
Vela remnant (Aschenbach 1998; Iyudin et al. 1998).  
Aschenbach (1998) found the object, RX J0852.0--4622
(hereafter abbreviated to RX J0852), in the {\it ROSAT}
image of Vela by restricting the photon energies to those
$> 1.3$ keV.
An  shell source then appears.
There is some emission from
other parts of Vela, including one of the bullets (Aschenbach et al. 1995),
but it is less prominent.
Iyudin et al. (1998) independently found evidence for a young remnant
in this region by the detection of $^{44}$Ti emission, making it the
second young remnant observed in this line after Cas A.
The strength of the emission, together with the remnant properties,
require a distance of $\sim 200$ pc and an age of $\sim 680$ years
(Iyudin et al. 1998; Aschenbach, Iyudin, \& Sch\"onfelder 1999).
However, Sch\"onfelder et al. (2000) have recently reanalyzed the
{\it COMPTEL} data on RX J0852 and found that the results depend on
the method of analysis.
They conclude that the evidence for a $^{44}$Ti signal from RX J0852
is at the $2\sigma - 4\sigma$ significance level and that an independent
discovery of a $^{44}$Ti supernova remnant cannot be claimed.
In view of these uncertainties, here we investigate a model where
RX J0852 is not a young remnant in its own right, but is a part
of the Vela remnant, perhaps 
created by a fast moving clump of ejecta.

The well-defined protrusions around the Vela remnant are suitable for
hydrodynamic modeling and that is the main subject of this paper.
The aims are to elucidate the properties of the interaction and to
place constraints on the properties of the ejecta clumps.
The work is related to that of Wang \& Chevalier (2001; hereafter 
WC01) on instabilities
and clumping in Type Ia SNRs, especially Tycho's remnant.
The differences here are the supernova profile appropriate to a core
collapse event, the greater age of Vela, and presence of well-defined
protrusions.
In \S~2, we present our computational setup and the method.
The evolution of nonlinear inhomogeneities is given
in \S~3.
The application of these results to the protrusions in Vela is in \S~4.
In \S~5, we discuss the possibility that the apparent SNR RX J0852
is related to clump interaction with the forward shock front
of the Vela remnant and is
not a separate SNR.
Our conclusions and discussion of future propects are in \S~6.

\section{SUPERNOVA DENSITY STRUCTURE  AND METHOD}

The existence of the Vela pulsar near the Vela remnant's center implies
a supernova origin  in gravitational core collapse.
For such supernovae, a strong shock wave passes
through the star, leading to two density components:
an inner flat portion and an outer power law profile 
(e.g., Chevalier \&  Liang 1989):
\begin{equation}
 \rho_{SN} =   \left\{  \begin{array}
                     {l@{\quad\quad}l}
                     F t^{-3}              &  v < v_{tr} \\
                     F t^{-3}(v/v_{tr})^{-n} &  v > v_{tr}
                     \end{array} \right.
\end{equation}
where
\begin{equation}
F={1\over 4 \pi n} { [  3(n-3)M ]^{5/2} \over 
                     [ 10(n-5)E ]^{3/2}  },
\end{equation}
\begin{equation}
v_{tr} =   \left[{10(n-5)E \over 3 (n-3)M }\right]^{1/2},
\end{equation}
$v_{tr}$ is the transition velocity  between the two density
components, and  $v=r/t$ is the flow velocity.  
The initial supernova explosion powers a spherical expansion of the ejecta so
that a gas element expands freely with a constant
velocity and its density drops as $t^{-3}$.
The ejecta density structure is determined by three parameters: 
the density power index $n$, the supernova's total explosion 
energy $E$, and the total ejecta
mass $M$.  
The power law index $n$ is generally taken to be in the range $7-12$.
For Vela, we take   $n=8$;
the flat component then comprises 5/8 of the
total ejecta by mass. 
A better representation of the density distribution would be to allow
for a continuous change from the inner flat profile to the outer
power law part, as expected in a realistic situation.
Matzner \& McKee (1999) have made calculations of such profiles for
core collapse supernovae.
However, we do not expect that the our results on clumps are substantially
affected by the details of the background density profile.

The interaction between an outer power law ejecta profile with $n > 5$ 
and a constant ambient density $\rho_{am}$    
 gives rise to an intershock solution 
that evolves in a self-similar way (Chevalier 1982).
With $n=8$ ejecta,
the characteristic radii of the intershock structure evolve as
\begin{equation}
R = A \left({E^5\over {\rho_{am}}^2 M^3} \right)^{1/16} t^{5/8},
\end{equation}
where $A$ is 1.08, 0.94, and 0.89 for the forward shock, the contact
discontinuity, and the reverse shock, respectively.  
The self-similar phase ends when the inner flat ejecta run into
the reverse shock
at the  time
\begin{equation}
t_c= 0.29 \left({M^5 \over {\rho_{am}}^2 E^3} \right)^{1/6}.   
\end{equation}
The reverse shock then starts to propagate back
to the explosion center relative to the self-similar expansion.

The computational methods in this paper are similar to that of our previous
study on Type Ia SNRs (WC01). 
We place density inhomogeneities in the unshocked,
freely expanding ejecta to track
the clump-remnant interactions at later times.                        
 
To acquire the one-dimensional intershock solutions,
we   initiated the computation with
  a freely-expanding power law density profile $n=8$ 
representing the ejecta on the inner side of the grid,
and a stationary ambient medium on the outer side. 
The inner numerical boundary condition was changed
 to accomodate the transition to
 flat ejecta  after the time $t_c$.
The one-dimensional intershock profile was then used to initiate two-dimensional simulations.
The grid was radially expanding, following the intershock boundaries.
We neglected the effects of magnetism, heat conduction, and radiation. 
The gas pressure in the unshocked gas 
is not important compared to the kinetic energy.
The filamentary optical emission from Vela indicates that there
are radiative shock waves in the remnant.
These  probably represent cloud interaction, as there is evidence for a number
of density components in the shocked gas, but much of the
shock interaction may be nonradiative (see \S~4).
Our assumption of no radiation is a first approximation to the evolution.
We used an adiabatic index $\gamma={5/3}$.  

Although our simulations are for a particular set of parameters,
through scaling we can obtain  solutions corresponding 
to various explosion parameters and ambient densities.  
The scaling parameters follow those of Dwarkadas \& Chevalier
(1998)  and WC01 (see also Truelove
\& McKee 1999):
we define $R'=(3M/4\pi\rho_{am})^{1/3}$, $V'=(2E/M)^{1/2}$, and $T'=R'/V'$, to  
describe the solution in the nondimensional quantities $r'$, $t'$, and $v'$, where
$r'=r/R'$, $t'=t/T'$, and $v'=v/V'$.  
For example, for an ejected mass of $10\Msun$,
 an explosion energy of $10^{51}$ ergs, 
and an ambient density of $2.34\times 10^{-24} \gcm$
(corresponding to $n_H=1$ cm$^{-3}$ with a H/He ratio of 10/1 by number)
along with $n=8$,
$R'= 4.1 $ pc, $V'=v_{tr}=3162 \kms$, and $T'= 1271$ yr. 
The nondimensional solutions can be re-scaled back to the dimensional solutions
corresponding to a different set of explosion conditions;           
the nondimensional solution sequence thus
represents all possible dimensional solutions with varying sets of initial 
parameters. 

\section{EVOLUTION OF CLUMPS}\label{sec:clump}

\subsection{Numerical Simulations}

One-dimensional simulations illustrate the overall hydrodynamic
situation in which the clumps are placed.
Fig. 1 shows the evolution of characteristic radii for 
our one-dimensional simulations.
The self-similar phase occurs for $t' \la 0.7$, 
when both the forward and reverse shock move 
steadily with a constant expansion parameter
$m=0.625$, where $m=(dR/dt)/(R/t)$.  
At $t' \approx 2.5$   
the reverse shock starts to turn over in a fixed frame.
At $t' =5.7$, the reverse shock reaches the explosion center.
The expansion approximately reaches the Sedov 
phase at $t' \approx 10$.

In the absence of clumps, our two dimensional
 simulations show the Rayleigh-Taylor
instabilities that are expected when the ejecta are decelerated by
their interaction with the surrounding medium.
These instabilities have been well described in previous 
numerical studies
(Chevalier, Blondin, \& Emmering 1992; Jun \& Norman 1996; Kane,
Drake, \& Remington 1999;
WC01).
One aspect of the present work is the inclusion of a flat portion
of the density profile.
When the reverse shock front enters this part of the density profile,
the instabilities begin to decay, as in the case of an exponential density
profile (WC01).
The unstable flow provides a background for our simulations of
clump interaction.
Fig. 2 shows angle-averaged radial density profiles superposed on
the one-dimensional density profiles at 4 times;
the first 3 times are ones at which the clump passes the reverse shock
front in  our simulations.

In the simulations, we aim to reproduce the initial   
conditions for Vela's bullets, although our results should also be
applicable to other supernovae.
Our method is similar to that used in our discussion of Type Ia
supernovae (WC01).
The differences here are the density profile for a core collapse supernova,
the need for significant protrusions in the outer shock front,
and the greater age of the remnant.
We placed dense regions in the ejecta to represent clumps 
in order to 
examine their hydrodynamical
 interaction with the remnant.
The clumps were initiated at the polar angle $\theta = 45 ^\circ$.
Although the nonzero polar angle makes a toroidal clump in our
two-dimensional simulations, 
the dynamics of such a cloud are likely to be similar to those  
of a spherical cloud (WC01). 
In our simulations, the grid lines were set linearly expanding from 
the inner grid boundary so that the zone spacing was kept uniform with time.
We used 1/2 of a quadrant or $\sim$1 quadrant centered at $\theta=45^\circ$.  
To prevent  
the reverse shock from being reflected from the side boundaries 
as it turned over, 
we applied periodic boundary conditions to these boundaries. 

The global evolution can be determined by three parameters: 
the initial density contrast, $\chi $, between the clump and supernova ejecta;
the initial impact time with the reverse shock, $t_0'$;
and the initial size of the clump.
We typically took $\alpha_0=1/3$, where 
$\alpha_0$ is the ratio of clump size to the intershock width 
between the forward and reverse shocks, at the time that the
clump reaches the reverse shock front.
Table 1 gives the corresponding ratio
of the clump size to its radius.  
The deceleration of a shocked clump is determined by the drag of the 
surrounding
material.
A denser and larger clump   
 is more able to resist crushing.
Once the shocked clump laterally expands into a crescent shape, 
its cross sectional area and drag are increased, and it becomes 
significantly decelerated. 
In the cases of Type Ia SNRs (WC01), 
clumps with $\alpha_0 \le 1/3$ are found to require $\chi \ga 100$                 
to survive crushing and 
cause a significant protrusion on the forward shock.
Here we similarly examine the initial density inhomogeneity for 
Vela's bullets.
We initiated clump interactions at various dynamical epochs, 
with $\chi$ up to 1000 and $\alpha_0$ up to 1/3 (Table 1).

Figs. 3 and 4
show the interaction with a single clump for $\chi=100$ and 
$\chi=1000$ with $\alpha_0=1/3$.
The basic physics of the interaction is related to that for the interaction
of planar nonradiative blast waves with an interstellar cloud 
(Klein, McKee, \& Collela 1994).
As the clump impacts on the reverse shock, a
transmitted `cloud shock' 
is driven into the clump and crushes it.
A larger density contrast between the clump and the interclump medium
causes a larger velocity 
difference, helping the development of a shear flow
and the Kelvin-Helmholtz instability
at the cloud-intercloud interface.
When the cloud shock exits the cloud, a rarefaction
wave moves back into the clump and causes lateral expansion. 
The cloud
gradually becomes flattened
and curved like a crescent.
Material streams out from the horns
of the crescent; the ram pressure difference between the axis
and the side of the cloud drives the mass loss.
The  pressure near the cloud axis was higher because of the additional
ram pressure on the front face of the clump. At the rear of the clump
the flow became turbulent and
left a trail of vorticies.
The acceleration
leads to the Rayleigh-Taylor instability on the upstream side of the clump.
The combined
 instabilities lead to the destruction of the clump.

The clump-remnant interaction has two regimes based on the intershock 
structures: 
the steep power law regime, when the shocked
region is self-similar, and  the flat density regime.
During the power law regime, 
the fragmentation process does not depend on the initial dynamical age, 
because the 
profiles of the physical variables
remain  constant.
For a given clump size/radius ratio and $\chi$, the evolution can be
scaled to other radii and times.
During the flat density regime, 
the ratio of the density at the reverse shock to that at
the forward shock decreases (Fig. 2), 
and clumps must traverse a relatively larger column density of matter
to reach the forward shock front,
so that a clump with a given value of $\chi$ is more effectively
stopped by the swept up surrounding medium.
Clumps initiated earlier are thus able to cause
protrusions in the forward shock 
with a smaller density contrast. 
This is similar to the case of Type Ia SNRs (WC01).

For the cases listed in Table 1 with $\chi=100$, we found that
({\it a}) in the power law regime for $t_0'=0.004$ and $t_0'=0.028$,
the forward shock restores spherical symmetry by $t'=0.3$;   
({\it b}) for $t_0'=0.8$, as the reverse shock has just entered the flat 
ejecta, 
the protrusion reaches its maximum strength at $t'=4.0$,
with an extent of $< 30\%$ in radius;
({\it c}) for $t_0'=1.4$, the protrusion is less than 20\%;
({\it d}) for $t_0'=2.22$ and $t_0'=4.1$,
the clump is quickly destroyed and
does not affect the forward shock.
 At this stage, the reverse shock has started propagating back to the explosion
 center.

When the clump initial density was increased to $\chi=1000$,  
clumps initiated early in the power law regime remain 
ineffective in causing a protrusion after $t' \approx 1$.
The $t_0'=0.8$ clump produces the strongest protrusion 
at $t'\approx 5-10$.
Although the remnant outline can also be disturbed in other cases 
over a large expansion factor, 
the $t_0'=0.8$ case can best cause a protrusion lasting to 
 times $t'\ga 5$.
To reproduce a 40\% bulge at the present epoch in
the Vela remnant, presumably at $t'\approx 5$ (see \S~4),
the initial clump-reverse shock impact should take place 
close to the transition between the power law regime
and the flat regime.
The reverse shock wave encounters the transition point
at $t'=0.7$.

One property of the flow after the clump moves ahead of the
forward shock front is that the pressure is low immediately downstream
from the clump.
In our model with $t_0'=0.028$ (or other cases in the self-similar
phase) and $\chi=1000$,
the pressure ratio between the head of the clump and the minimum pressure
downstream region is $20-300$ for $t'$ in the range $0.05-0.5$.
As the clump is broken up, the low pressure region moves downstream
from the clump and occurs in parts of the flow with a large vortical motion.
The low pressure regions also have a low density.

We have compared our lower resolution runs with 300 by 300 zones
to runs with 600 by 1000 zones (Fig. 5).
At higher resolution, more structure appears in the flow, but the
larger scale results and the density contrast that we determined
are not changed.
Fig. 5 shows that there is a narrow protuberance at the leading
part of the flow at higher resolution, but it does not ultimately lead to 
a substantially more extended protrusion.
X-ray and radio observations of supernova remnants
may eventually be able to delineate the outer
shock front for comparison with simulations, although the
unknown structure of the clump could be a factor in the result.

\subsection{Interpretation}

In the case of a cloud that is passed over by a blast wave
(Klein et al. 1994), the destruction of the cloud occurs on
several cloud crushing times.
The destruction time is approximately the time at which we
expect strong deceleration of a clump.
The velocity of the shock in the cloud is determined by the
surrounding pressure, which is $\sim \rho_{am} v_b^2$, where
$v_b$ is the velocity of the blast wave.
In the present case, when the clump has extended beyond the
forward shock wave radius, the ram pressure at the head of the clump
is $\sim \rho_{am} v_c^2$, where
$v_c$ is the clump velocity in the fixed frame and $\rho_{am}$ is the
density outside the forward shock wave.
Before the clump reaches the outer shock, it is in the hot
shocked shell and the total pressure experienced by the
clump is given by the local value of $p+\rho v_{rel}^2$,
where $v_{rel}$ is the relative velocity between the clump
and the surrounding medium.
During the early self-similar phase, both the forward and reverse
shock radii increase as $t^m$, where $m=5/8$ for $n=8$.
Using the postshock conditions at the reverse shock ($p=
0.75\rho_r [1-m]^2 R^2/t^2$, $\rho=4\rho_r$, $v_{rel}=0.75[1-m]R/t$), we
find $p+\rho v_{rel}^2 = 3\rho_r (1-m)^2 R^2/t^2$, where
$\rho_r$ is the preshock density at the reverse shock and $R$ is
the reverse shock radius.
The value of $p+\rho v_{rel}^2 = 0.42\rho_r  R^2/t^2$ for $n=8$
stays fairly constant in the shocked region (Chevalier 1982).
For $n=8$, $\rho_r/ \rho_{am}=2.1$ (Table 1 of Chevalier 1982)
 and the clump is freely expanding
so that $v_c=R_1/t_1$, where $R_1$ is the reverse shock radius
when the clump reaches the reverse shock at time $t_1$.
Then clump then initially experiences a total pressure of $0.9\rho_{am} v_c^2$.
These considerations show that the clump experiences a fairly
even total pressure after it moves through the reverse shock.

Another property of the clumps in our case
is that they are initially uniformly expanding,
so that their density decreases as $t^{-3}$.
If the clumps travel a significant radial distance, this change is
important.
If a clump reaches the reverse shock front (radius $R_1$)
 at time $t_1$ and it continues to
move at a constant velocity, it reaches the forward shock (radius $1.21 R_1$
moving as $t^{5/8}$ for $n=8$) at time $1.66 t_1$.
There is some evolution of the density and the properties of the
shocked shell during this time.

We now consider an approximate model for the crushing of a clump.
A clump moving with velocity $v_c$ reaches the reverse shock front
at time $t_1$ when its size is $a_1$ and density is $\rho_1$.
If the cloud is crushed rapidly compared to $t_1$, the crushing time
is $t_{c1} \approx a_1 v_c^{-1}(\rho_1 /\rho_{am})^{1/2}$.
If the clump is more long-lived, the variation of the clump size, $a$,
in the frame of the clump can be estimated from 
the equation of motion for the
shock position in the expanding clump:
\begin{equation}
{da\over dt} = {a\over t} -v_c\left(\rho_{am}\over \rho_1\right)^{1/2}
\left(t\over t_1\right)^{3/2},
\end{equation}
with the solution
\begin{equation}
a=a_1{t\over t_1} - 
{2v_c\over 3}\left(\rho_{am}\over \rho_1\right)^{1/2}{(t^{3/2}
-t_1^{3/2})t\over t_1^{3/2}}.
\end{equation}
The shock is initially carried out by the clump expansion, but the
decrease in density eventually leads to a relatively rapid compression of the
clump.
The clump is crushed ($a=0$) at time
\begin{equation}
t_{cr}=t_1\left(1+{3t_{c1}\over 2t_1}\right)^{2/3},
\end{equation}
where $t_{cr}$ is now measured from $t=0$.
We have $\rho_1 /\rho_{am} =2.1\chi$ during the self-similar phase,
so $t_{cr}\propto a_1^{2/3} \chi^{1/3}$ when
$3t_{c1} >2 t_1$.
For the self-similar case with $a_1/R_1=0.075$, we find
$t_{cr}\approx t_1(1+0.16\chi^{1/2})^{2/3}$; for $\chi=1000$, $t_{cr}=3.3t_1$.
Once the clump is crushed, the initial expansion is no longer a factor
and we expect clump ``destruction'' on several crushing times, as
in Klein et al. (1994).  This is in accord with the stopping times
noted above.
Denser clumps are expected to be more robust given the same mass,
because the initial cloud crushing time is  
$t_{c1}\approx {\chi}^{1/2} a_1/v_{c}$ and $\chi\propto a_1^{-3}$ for 
a fixed mass, so that $t_{c1}\propto \chi^{1/6}$.

\section{THE VELA SUPERNOVA REMNANT}  \label{sec:discussion}

The Vela  SNR is a middle-aged Galactic SNR.
Its overall emission shows an extended circular limb-brightened
shell similar to the Cygnus Loop, and a
 pulsar-powered nebula Vela X
like the Crab Nebula.
We take the age to be the
characteristic spin-down time of
the Vela pulsar PSR B0833-45 assuming magnetic dipole
radiation, $ 11400$ yrs
(Reichley, Downs \& Morris 1970).
Lyne et al. (1996) have measured a braking index of $n=1.4\pm 0.2$
for the pulsar,
which would suggest a larger age, but the result is uncertain.
Recent studies suggest a distance of 250 pc (Cha, Sembach \& Danks 1999),
placing Vela the closest SNR to Earth.
The observed diameter of $8^\circ$ (Aschenbach et al. 1995) gives
it a radius of 17 pc.

There are the 
six X-ray bullets or knots (Aschenbach et al. 1995) protruding 
beyond the  remnant boundary by up to 40\% of the blast wave radius.
Of these, 5 have been detected at radio wavelengths (Strom et al. 1995;
Duncan et al. 1996).
In the radio, knot A, which lies $5.3^{\circ}$ from PSR 0833-45, 
is especially prominent and shows a greater extent than at X-ray wavelengths.
Knots A and E extend 
beyond the remnant edge by $1.2^\circ$ and $2.4^\circ$, respectively.
The trailing Mach cones indicate that the knots 
have moved supersonically through their surroundings.
The symmetry axes of these knots intersect
 close to the remnant's geometric center, suggesting an association 
 with the supernova, rather
than with density inhomogeneities of the surrounding medium.  
The finding of a Si overabundance in knot A (Tsunemi et al. 1999)
and O, Mg, and Si overabundances in knot D (Slane et al. 2001b),
 as deduced from X-ray
observations, also provides strong support for an ejecta clump origin
for these features.

Recent {\it Chandra} observations show 
that the pressure in the head region of knot A 
is $\sim 10$ times higher than that in the tail region
 (Miyata et al.  2001).
As discussed in the previous section, a pressure differential 
of this magnitude
is expected to be present between the head and downstream region
of a fast knot.
The morphology of knot A  and its trail indicates that it is still in an
early expansion phase and has not yet greatly decelerated and
spread laterally.
The tail region observed by Miyata et al. (2001) is immediately
downstream from the bright head region.
This is the region where is low pressure is expected, before the
strong deceleration.
The pressure observations thus support a picture with fast clumps,
as opposed to structure formed by an inhomogeneous blast wave.

A limitation of our models is that we have assumed a constant
density surrounding medium, although there is evidence that
the surroundings of Vela are inhomogeneous.
Different densities have been deduced from observations at different
wavelength regions.
From X-ray observations, where hot low density gas is observed, 
Bocchino et al. (1999) claim the presence of an ambient intercloud
medium with $n_0=0.03$ cm$^{-3}$ and denser inhomogeneities with
$n_0=0.13$ cm$^{-3}$.
Raymond et al. (1997) have studied a face-on radiative shock wave
in the remnant at optical and ultraviolet wavelengths, finding a shock
velocity of $170\kms$ and an ambient density of $n_0=2.7$ cm$^{-3}$.
H I observations by Dubner et al. (1998) indicate the presence of
a partial shell of neutral gas moving at $30\kms$; they deduce
a preshock hydrogen density of $n_0\approx 1-2$ cm$^{-3}$.
Dubner et al. (1998) argue that the neutral gas makes up a shell of
material, but the multiwavelength observations and the low
velocity suggest that this
represents the shock front moving into clouds.

Vela is the result of a core collapse supernova, so that it
is likely that the progenitor star was massive and affected its
environment through photoionization and stellar winds
(e.g., Gvaramadze 1999).
It is possible that the partial H I shell is related to such
effects.
However, the assumption of a constant density surroundings with density
$0.1-0.2$ cm$^{-3}$ provides a first approximation for the
dynamics.
Because of the appearance of the Vela X pulsar nebula and the
displacement of the pulsar from the nebula, Blondin, Chevalier,
\& Frierson (2001)
argued that the reverse shock front in Vela has returned to the
center and affected the pulsar nebula.
From Fig. 1, this requires $t^{\prime} \ga 5.7$ for Vela.
At this phase, the remnant is not well into the Sedov blast wave
regime, so in Fig. 6 we show the required ambient density
and explosion mass in order to obtain a particular value of $t^{\prime}$,
for an energy of $10^{51}$ 
and the current radius of Vela.
For $n_0=0.1$ cm$^{-3}$, $E=10^{51}$ ergs, and $M=8\Msun$, the
reference quantities discussed in \S~2 are
$R'= 8.2 $ pc, $V'=v_{tr}=3,540 \kms$, and $T'= 2,280$ yr. 
The age is $t^{\prime}\approx 5$.
The fact that the protrusions extend to $\sim 40$\% of the remnant
radius and the considerations of the previous section imply
 clump velocities in the freely expanding ejecta of
$\sim 3,000\kms$.

Our models assume that radiative cooling is not important
for the evolution.
The pressure-driven snowplow, or radiative cooling, phase is
expected to begin at a radius (e.g., Cioffi et al. 1988)
\begin{equation}
R_{PDS}= 37\left(E\over 10^{51}{\rm~ergs}\right)^{2/7}
\left(n_0 \over 0.1{\rm~cm}^{-3}\right)^{-3/7} {\rm~pc},
\end{equation}
which is consistent with Vela currently being in the nonradiative phase.
However, radiative shocks are present where the remnant in running
into denser interstellar gas.
The protrusion `D' appears to be driving a radiative shock wave
into the surrounding medium on one side (Redman et al. 2000),
which we attribute to an interstellar inhomogeneity.

The density in the inner, flat part of the supernova density profile
can be obtained from equation (1).
The density of a clump in this region is
\begin{equation}
\rho_{cl}=6\times 10^{-24}
\left(\chi\over 10^3\right)
\left(M\over 8\Msun\right)^{5/2}
\left(E\over 10^{51}{\rm~ergs}\right)^{-3/2}
\left(t\over 10^{4}{\rm~yr}\right)^{-3} {\rm~g~cm}^{-3}.
\end{equation}
An estimate of the cooling rate at the present time, taking into
account O-rich gas with a temperature of $(3-6)\times 10^6$ K
(Aschenbach et al. 1995; Tsunemi et al. 1999) yields a cooling time
somewhat larger than the age using the equilibrium cooling curve
for an O gas (Borkowski \& Shull 1990).
This is consistent with the lack of optical emission that might be
associated with heavy element gas.
However, the cooling for a Si gas is more rapid (Hamilton \& Shull
1984) and at an earlier stage of evolution when the protrusion is growing,
the density is higher and radiative effects may be important.
A complete treatment of radiative cooling of a heavy element gas
is outside the scope of the present paper.

In our simulations, we have taken a low sound speed in the ambient
medium, so that the motion of the clumps is highly supersonic.
The results do not depend on the ambient sound speed provided
the  clump velocity is at least moderately supersonic.
Aschenbach et al. (1995) have estimated relatively low Mach numbers for
the clumps (2.4--4.0) based on the opening angles of the observed
protrusions.
However, those estimates assume that the clump trail is formed
by a clump moving at a constant velocity.
We expect that the clumps  initially move out rapidly, but then
are decelerated by their interaction with the surrounding medium.
Since they spend most of their time in the decelerated state,
most of the observed knots are probably strongly decelerated so that
the opening angle cannot be used to determine the Mach number.

\section{IS RX J0852 A SEPARATE SUPERNOVA REMNANT?}

The {\it ROSAT} source RX J0852 is a shell X-ray source
that appears in harder X-rays superposed on the Vela remnant
(Aschenbach 1998).
Here, we investigate the possibility that it is part of the Vela remnant.
We have found that as an ejecta clump makes a protrusion in the
supernova remnant shell,
the ram pressure of the external medium causes instabilities and
lateral spreading of the clump.
The clump decelerates and the rest of the blast wave catches up with it.
The deceleration of the clump effectively deposits the kinetic energy
of the clump into the supernova shell.
The growth of the disturbance in the main shell can be seen in Figs. 3 and 4,
which show that the outer edge of the disturbance is accompanied
by a region of somewhat higher density.
This region has a sharp edge, which is a tangential discontinuity.

The required properties of the clump can be estimated from observations
of RX J0852 and the Vela remnant.
The angular diameter of RX J0852 is about $2^{\circ}$, as compared to
the $8^{\circ}$ diameter of the Vela remnant.
The outer diameter of the corresponding radio features is
$1.8^{\circ}\pm 0.2^{\circ}$ (Duncan \& Green 2000).
The RX J0852 source thus covers  $\sim 10^{-2}$ of the surface of the
supernova remnant.
Assuming a supernova energy of $10^{51}$ ergs for Vela, the energy in the
affected solid angle is $10^{49}$ ergs.
We express the energy deposited by the fast clump as $10^{49}\beta$ ergs
where $\beta <1$ because there is little pressure enhancement in
the disturbed region at the times of interest; in fact, $\beta < 0.1$
is likely.
The clump interaction is in a more advanced state than that
of the current protrusions in Vela, for which we estimated
an initial velocity of $\sim 3,000\kms$.
If we take the same velocity here,
the estimated clump mass is  $0.1\beta\Msun$.
The energy and mass  are similar to those estimated for the knots
in Vela and their trails (Aschenbach et al. 1995).

The thermal energy in RX J0852 has been estimated from the X-ray emission.
Chen \& Gehrels (1999) and Aschenbach et al. (1999)
made estimates of the energetics of RX J0852 based on the assumption
that it is a separate supernova remnant and is a $^{44}$Ti source.
They found that the estimated energy is low.
For their preferred values, Aschenbach et al. (1999) find an energy
of $2.6\times 10^{49}$ ergs for a distance of 200 pc, although
the uncertainty is large.
The fact that the X-ray spectrum of RX J0852 appears to be
nonthermal (Allen et al. 1999; Slane et al. 2001a)
 means that the X-ray flux actually gives an upper
limit on the thermal emission.
The observed X-ray spectrum is $F_{\nu}\propto \nu^{-1.6}$
(Slane et al. 2001a), which corresponds to the steep particle spectrum
$N(E)\propto E^{-4.2}$ for synchrotron emission.
If the X-ray emission is due to the steeply declining tail of
the electron distribution, a relatively small disturbance in the
total energy density may be able to give the observed effect.
The lack of the shell in the complete {\it ROSAT} band indicates that
there is not a large density jump.
We speculate that the larger shock velocity in the region of the
protrusion enhanced the production of high energy electrons and
the observed shell is the remnant of that effect.

The morphology of the source edge in our model is a toroidal region
on the surface of the Vela remnant.
In view of the fact that RX J0852 is near the edge of the remnant,
foreshortening of the ring would be expected.
The observed shell is not clearly present on the East side, so this
is difficult to test.
The observed source does show some elongation in the expected
North--South direction.

A possible problem for our model is that if RX J0852 is the result of
a clump interaction, similar emission regions might be created by
other clumps.
Yet, the $>1.3$ keV image of Aschenbach (1998) does not show other
features similar to RX J0852.
In our model, the clump does not create a shell feature when it
is causing a protrusion, but Figs. 3 and 4 show that as a clump that
has caused a substantial protrusion moves back toward the main
shell, there is a density disturbance at the intersection of the
protruding shock with the main shell.
Once a clump has decelerated and expanded laterally, the enhanced
density ring
generated in the shell is a transient feature that is expected to
weaken with age, so that features of this type may be short-lived.
Allen et al. (1999) find that both RX J0852 and the Vela remnant have
similar spectra in the $3-12$ keV range; it is difficult to
distinguish them, providing a further connection between RX J0852
and the Vela remnant.

The radio emission associated with RX J0852 has been discussed by
Combi, Romero, \& Benaglia (1999) and Duncan \& Green (2000).
Duncan \& Green (2000) give a more detailed discussion and mention
several aspects of RX J0852 that make it plausible that it is part
of the Vela remnant:
1)  The surface brightness of RX J0852 is similar to that of the
Vela remnant.
Before the finding of the X-ray shell, the object was assumed to be
part of the Vela remnant.
If the remnant is as close as 200 pc, its radio surface brightness is
unusually low for a supernova remnant, even compared to SN 1006.
2)  Radio features of the remnant appear to join up with those
in the Vela supernova remnant.
In particular, the eastern edge continues to the North in a filament
that is clearly part of the Vela remnant.
Extensions that Combi et al. (1999) suggested may be part of RX J0852
can be identified as structure in the Vela remnant.
3)  The radio spectral index of the bright North part of RX J0852
is estimated to be $\alpha=-0.4\pm 0.15$, where flux 
$F_{\nu}\propto \nu^{\alpha}$.
This value is unusually flat for a young supernova remnant, but is
comparable to that found in other parts of Vela's supernova remnant
shell (Dwarakanath 1991).
4)  The values of rotation measure and magnetic field direction deduced
from the radio polarization do not show any discontinuities across the
shell of RX J0852.
If RX J0852 is a separate object, it does not contribute to the
polarization properties in this region of the Vela remnant.

The model proposed here for RX J0852 makes a number of testable
predictions.
One is that it should not be a source of $^{44}$Ti emission at the level that
was initially claimed (Iyudin et al. 1998).
This can be tested by observations with the forthcoming {\it INTEGRAL}
mission.
Another is that X-ray emission from the disrupted clump may be detectable
in  RX J0852.
Tsunemi et al. (2000) found evidence for a Ca overabundance on one
side of RX J0852 from X-ray observation with {\it ASCA}.
They attribute the presence of Ca to the decays of $^{44}$Ti, but
it is possible that it comes from ejecta clumps.
In addition, they note that the lower energy emission from the region
of RX J0852 is very much like that from the Vela remnant and they
suggest that Vela is the source of the emission.
From {\it ASCA} observations,
Slane et al. (2001a) found evidence for diffuse central emission, but also
for a possible compact X-ray source, which is not expected in our model.
Mereghetti (2001) has examined this region with {\it BeppoSAX} and
found that the candidate compact star is likely to be an early-type star,
but there is another unresolved source that is possibly a neutron star.
Pavlov et al. (2001) clarified the situation with {\it Chandra}
observations showing the presence of a compact X-ray source.
They suggest that the central source belongs to the newly
emerging class of radio-quiet young neutron stars.
If this identification is confirmed and the source is clearly linked
to RX J0852, the scenario discussed here must be rejected.

\section{DISCUSSION AND CONCLUSIONS}\label{sec:conclu}

Our simulations for the apparent bullets in the Vela remnant
show that they are likely to be ejecta clumps with an initial
velocity of $3,000-4,000\kms$ and a density contrast  
$\chi \ga 10^3$ compared to their surroundings.
Observations of core collapse supernovae show evidence for
such clumps in the heavy element ejecta at an age $\sim$ 1 year,
and we have assumed here that the clumps form at an early time
compared to the age of Vela so that the clumps are comoving with
the surrounding ejecta.
 
The origin of the clumps is uncertain.
The ejecta density structure observed in SN 1987A can be
attributed to the Ni bubble expansion effect (Li et al. 1993; Basko 1994).
One prediction of this model is that the expanding low density gas
is initially $^{56}$Ni, which decays to Fe.
It compresses the nonradioactive ejecta components, which is
consistent with the overabundance of Si observed in bullet `A'
(Tsunemi et al. 1999).
However, it is not clear whether the Ni bubble effect can give
the required compression ratio in the clumps.
Blondin, Borkowski, \& Reynolds (2001) studied the interaction of
supernova Ni--Fe bubbles with a surrounding medium.
They found deformation of the outer shock front, but not the type
of protrusions observed in the Vela remnant.
However, their calculations emphasized the low density bubbles
rather than high density regions in the bubble walls.
The computations of Basko (1994) show that the Ni bubble sweeps up
a shell with density compression up to a factor $\chi \sim 10$.
If radiative cooling is not important for the clumps, we find
that $\chi \sim 10^3$  is needed.
If cooling is important, there is less lateral expansion of the
clumps and a lower value of $\chi$ is presumably needed to obtain a given
protrusion.
  
Another mechanism that could operate in Vela is the sweeping
up of ejecta by a pulsar bubble, as is observed to occur in
the Crab Nebula.
In the Crab, the average density of the ejecta if they were
smoothed over the whole volume would be $n_H\sim 4$ cm$^{-3}$,
while the density in ionized filaments is $\sim 10^3$  cm$^{-3}$
(Davidson \& Fesen 1985).
The effective density contrast is thus $\sim 250$.
If there is cooler, neutral gas in a pulsar bubble, the compression
could be higher.
The existence of the Vela  pulsar and the Vela X nebula in the
remnant (Bock et al. 1998) shows that clumping by a pulsar
nebula is a possibility.

Our finding that clumpy ejecta can produce the protrusions
observed in the Vela remnant suggests that there may be
similar phenomena in the remnants of other core collapse
supernova remnants.
However, we have found that in remnants that are more evolved
than Vela, it is difficult to produce protrusions, so the
phenomenon is expected only in relatively young remnants.
There is widespread evidence for optically-emitting ejecta knots
in young core collapse supernova remnants, such as Cas A,
N132D, and Puppis A.
These are cases where no pulsar nebula is observed and
 where a complex interaction
with circumstellar gas 
appears to be taking place.
The finding of protrusions like those around the Vela remnant
may require deeper studies at X-ray and optical wavelengths.
 
In our picture, clumps move out ahead of the forward shock
front in the Vela remnant, are decelerated, and the forward
shock front catches up with them.
In this process, the kinetic energy of the clump is added to the
internal energy of the supernova remnant, which can result in
the formation of features in the outer part of the remnant.
We suggest that the apparent supernova remnant RX J0852-4622
superposed on the Vela remnant is structure in the Vela remnant
shell created in this way.
Current radio studies indicate that RX J0852 has properties
that appear to relate it to the Vela remnant.
Further X-ray studies should show whether its X-ray properties
clearly distinguish it from the rest of the Vela remnant.
Confirmation of either a $^{44}$Ti excess or a compact object
in RX J0852 would show that our conjecture is incorrect.
In addition, spatially resolved X-ray spectra, as
is possible with the {\it Chandra} and {\it XMM} observatories,
will be valuable for showing whether our picture of ejecta
clumps is valid.
We expect heavy element clumps to be fragmented by their interaction
with the surrounding medium.
 
We are grateful to J. Blondin and J. Hawley for help and comments
on the manuscript, and to R. Diehl for discussions of the
$\gamma$-ray emission from RX J0852--4622.
Support for this work was provided in part by NASA grant NAG5-8232.

\clearpage

 \begin{table}[h]

 \begin{center}
 \caption{\centerline{Clump Simulation Parameters} \label{tabclump}}
 \begin{tabular}{lllllll}
 \tableline
 \tableline
 $t'$     \tablenotemark{(1)}     & $\rho_c$ \tablenotemark{(2)} &
 $R_{rs}$ \tablenotemark{(3)}     & $R_{fs}$ \tablenotemark{(4)} &
  $v_{c}$  \tablenotemark{(5)} &
 $\alpha_{c}$  \tablenotemark{(6)} & 
 $M_{c}$  \tablenotemark{(7)}  \cr
 \tableline
 0.0041  &  3.086 & 0.0273 & 0.033  & 6.439  & 7.4\%  & $2.9
\times 10^{-5}$  \\
 0.028   &  2.916  & 0.0917 & 0.111  & 3.151 & 7.4\%  & $1.1
\times 10^{-3}$  \\
 0.80    &  1.226   & 0.738  & 0.901  & 0.886 & 7.6\% & $0.26$
\\
 1.40    &  0.229   & 0.958  & 1.291  & 0.638 & 12.4\% & $0.41$
\\
 2.22    &  0.0553  & 1.052 & 1.672   & 0.411 & 21.8\% & $0.69$
\\
 \tableline
 \end{tabular}
 \end{center}
 \tablenotetext{(1)}{starting time of the clump-reverse shock interaction}
 \tablenotetext{(2)}{clump density for $\chi=1$, normalized
 to the unshocked ISM}
 \tablenotetext{(3)}{radius of the reverse shock normalized to $R'$ }
 \tablenotetext{(4)}{radius of the forward shock normalized to $R'$}
 \tablenotetext{(5)}{initial clump velocity normalized to $V'=v_{tr}$ }
 \tablenotetext{(6)}{ratio of clump size to clump radial distance 
for $\alpha_0=1/3$}
 \tablenotetext{(7)}{clump mass ($M_{\odot}$) for $\chi=1000$ and
$\alpha_0=1/3$ with $M=10 M_{\odot}$ and $E=10^{51}$ ergs}
\end{table}

\clearpage


\figcaption[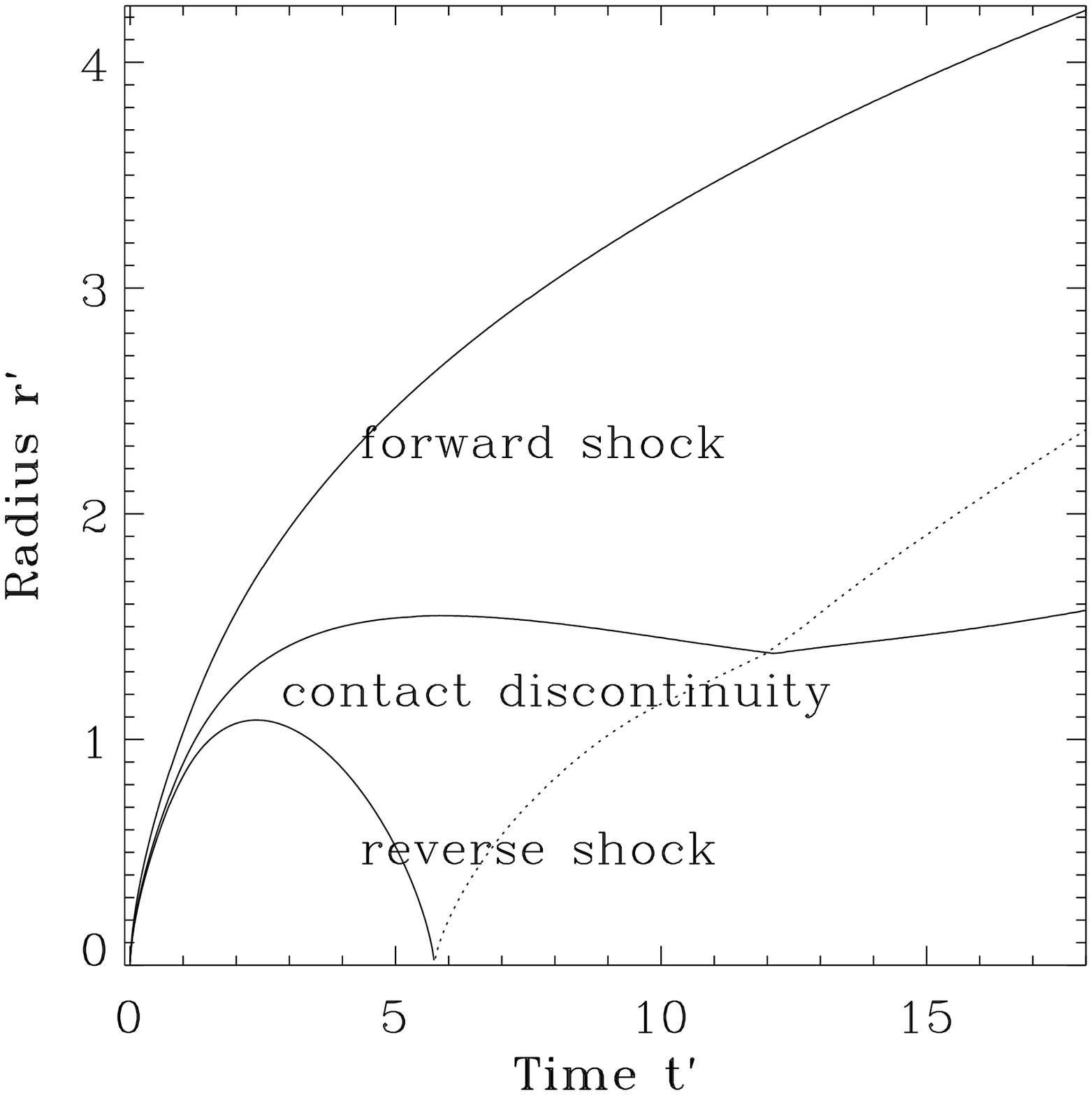]
{Evolution of the forward shock,
the contact discontinuity, and the reverse shock radius with time.
The dashed line shows the outgoing weak shock wave caused by the
reflection of the reverse shock wave at the center.
\label{fig.vela-radius}}


\figcaption[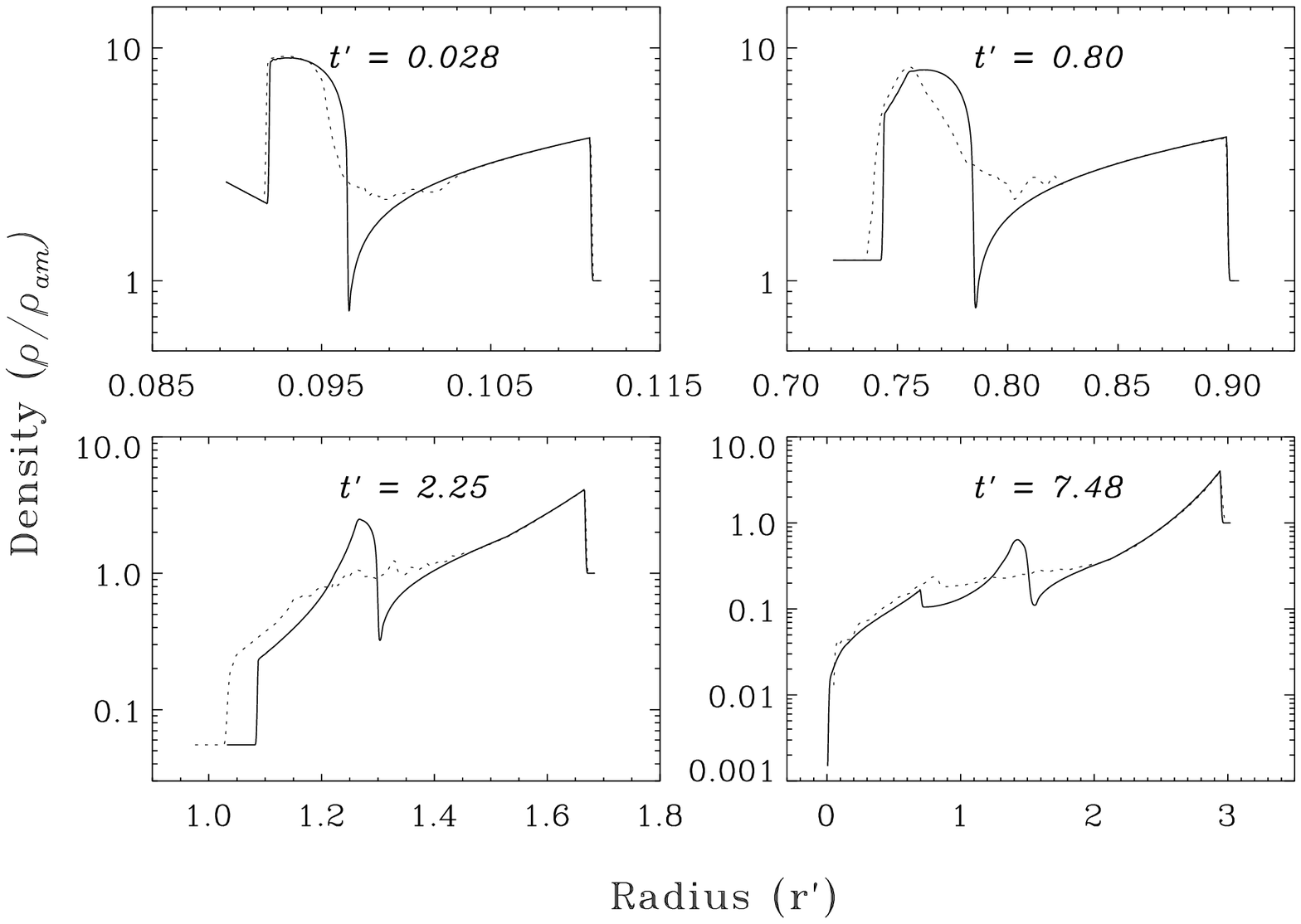]
{Angle-averaged two-dimensional density distribution  
plotted with the one-dimensional unperturbed solution at four 
stages of the evolution.
The time and radius use the normalized units given in the text.
The density is normalized to the ambient density.
In two dimensions, the reverse shock smears to
a smaller radius due to instabilities.
\label{fig.vela-dens1d}}

\figcaption[cl1-0.ps]
{Snapshots of a single clump in the ejecta expanding into the shocked region
starting in three dynamical stages
at $t_{0}'=0.028$, $t_{0}'=0.80$, and $t_{0}'=2.22$.
The clump has an initial density contrast $\chi=100$ to the
surrounding unshocked ejecta and a relative size $\alpha_0=1/3$ to the intershock
width.
The grid has 600 radial by 1000 angular zones centered on 
$\theta = 45^{\circ}$. The first two series (in columns) use 
1/2 of a quadrant and 
the third uses 8/9 of a quadrant.
\label{fig.cl1}}

\figcaption[cl2-0.ps]
{Snapshots of a single clump in the ejecta expanding into the shocked region
starting 
at the three times $t_{0}'=0.028$, $t_{0}'=0.80$, and $t_{0}'=2.22$,
with an initial density contrast $\chi=1000$ and a size $\alpha_0=1/3$.
The grid has the same resolution as in the previous plot.
\label{fig.cl2}}


\figcaption[cl3-0.ps]
{A high resolution clump-remnant simulation
initiated in the power law regime at $t_{0}'=0.028$ with $\chi=1000$
and $\alpha_0=1/3$ compared with a low resolution
one.
The high resolution  grid is 
600 radial by 1000 angular zones on 1/2 of a quadrant, 
and the low resolution one is 300 by 300 zones.
\label{fig.cl3}}

\figcaption[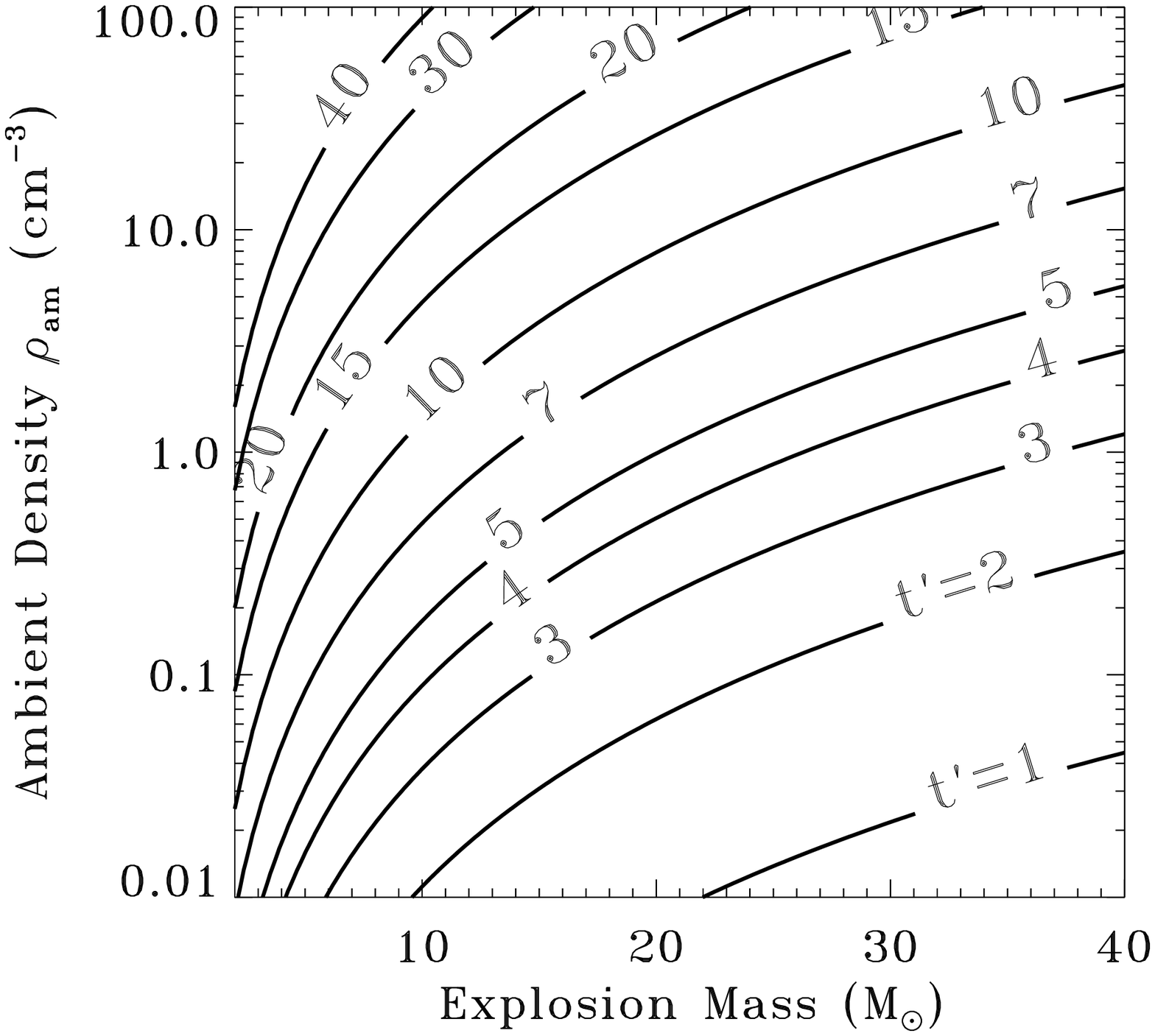]
{Variation of the  dynamical age of the Vela remnant (dimensionless time)
with the ambient density
and  explosion mass for an explosion energy of
$E=10^{51}$ ergs. 
The ambient density is in units of $2.34\times 10^{-24}$ gm cm$^{-3}$.
\label{fig.velaage}}


\clearpage



\begin{thebibliography} {}



\bibitem[]{}
Allen, G. E., Markwardt, C. B., \& Petre, R. 1999, BAAS, 31, 862

\bibitem[]{}
Aschenbach, B. 1998, Nature, 396, 141

\bibitem[]{}
Aschenbach, B., Egger, R., \& Tr\"umper, J. 1995, Nature, 373, 587


\bibitem[]{}
Aschenbach, B., Iyudin, A. F., \& Sch\"onfelder, V. 1999, \aap, 350, 997


\bibitem[]{}
Basko, M.  1994, \apj, 425, 264

\bibitem[]{}
Blondin, J. M., Borkowski, K. J., \& Reynolds, S. P. 2001, ApJ, 557, 782

\bibitem[]{}
Blondin, J. M., Chevalier, R. A., \& Frierson, D. M. 2001,  ApJ, in press
(astro-ph/0107076)

\bibitem[]{}
Bocchino, F., Maggio, A., \& Sciortino, S. 1999, \aap, 342, 839

\bibitem[]{}
Bock, D. C.-J., Turtle, A. J., \& Green, A. J. 1998, \apj, 116, 1886

\bibitem[]{}
Borkowski, K. J., \& Shull, J. M. 1990, \apj, 348, 169

\bibitem[]{}
Cha, A. N., Sembach, K. R., \& Danks, A. C. 1999, \apj, 515, L25

\bibitem[]{}
Chen, W., \& Gehrels, N. 1999, \apj, 514, L103

\bibitem[]{}
Chevalier, R. A. 1982, \apj, 258, 790


\bibitem[]{}
Chevalier, R. A., Blondin, J. M., \& Emmering, R. T. 1992, \apj, 392, 118

\bibitem[]{}
Chevalier, R. A., \& Kirshner, R. P. 1979, \apj, 233, 154

\bibitem[]{}
Chevalier, R. A., \& Liang, E. P. 1989, \apj, 344, 332

\bibitem[]{}
Cioffi, D. F., McKee, C. F., \& Bertschinger, E. 1988, \apj, 334, 252

\bibitem[]{}
Combi, J. A., Romero, G. E., \& Benaglia, P. 1999, \apj, 519, L177

\bibitem[]{}
Davidson, K., \& Fesen, R. A. 1985, ARA\&A, 23, 119


\bibitem[]{}
Dubner, G. M., Green, A. J., Goss, W. M., Bock, D. C.-J., \& Giacani, E.
1998, AJ, 116, 813


\bibitem[]{}
Duncan, A. R., \& Green, D. A. 2000, \aap, 364, 732

\bibitem[]{}
Duncan, A. R., Stewart, R. T., Haynes, R. F., \& Jones, K. L. 1996, MNRAS, 
280, 252


\bibitem[]{}
Dwarakanath, K. S. 1991, J. Astrophys. Astro., 12, 199

\bibitem[]{}
Dwarkadas, V., \& Chevalier, R. A. 1998, \apj, 497, 810

\bibitem[]{}
Filippenko, A. V., \& Sargent, W. L. W. 1989, \apj, 345, L43

\bibitem[]{}
Gvaramadze, V. 1999, \aap, 352, 712

\bibitem[]{}
Hamilton, A. J. S., \& Sarazin, C. L. 1984, \apj, 287, 282


\bibitem[]{}
Iyudin, A. F. et al. 1998, Nature, 396, 142


\bibitem[]{}
Jun, B.-I., \& Norman, M. L. 1996, \apj, 465, 800


\bibitem[]{}
Kane, J., Drake, R. P., \& Remington, B. A. 1999, \apj, 511, 335

\bibitem[]{}
Klein, R. I., McKee, C. F., \& Colella, P. 1994, \apj, 420, 213

\bibitem[]{}
Kundt, W. 1988, in Supernova Shells and their Birth Events, ed. W. Kundt
(Berlin: Springer), 245



\bibitem[]{}
Li, H., McCray, R., \& Sunyaev, R. A. 1993, \apj, 419, 824

\bibitem[]{}
Lyne, A. G., Pritchard, R. S., Graham-Smith, F., \& Camilo, F. 1996, 
Nature, 381, 497


\bibitem[]{}
Matheson, T., Filippenko, A. V., Ho, L. C., Barth, A. J., \& Leonard, D. C.
2000, \aj, 120, 1499

\bibitem[]{}
Matzner, C. D., \& McKee, C. F. 1999, \apj, 510, 379

\bibitem[]{}
Mereghetti, S. 2001, \apj, 548, L213





\bibitem[]{}
Miyata, E. M., Tsunemi, H., Aschenbach, B., \& Mori, K. 2001, \apj, 559, L45

\bibitem[]{}
Pavlov, G. G., Sanwal, D., Kiziltan, B., \& Garmire, G. P. 2001, 
\apj, 559, L131

\bibitem[]{}
Pavlov, G. G., Zavlin, V. E., \& Tr\"umper, J. 1999, \apj, 511, L45

\bibitem[]{}
Raymond, J. C., Blair, W. P., Long, K. S., Vancura, O., Richard, R. J.,
Morse, J., Hartigan, P., \& Sanders, W. T. 1997, \apj, 482, 881



\bibitem[]{}
Redman, M. P., Meaburn, J., O'Connor, J. A., Holloway, A. J., \& Bryce, M.
2000, \apj, 543, L153


\bibitem[]{}
Reichley, P. E., Downs, G. S., \& Morris, G. A., 1970, \apj, 159, 35


\bibitem[]{}
Sch\"onfelder, V., Bloemen, H., Collmar, W., Diehl, R., Hermsen, W.,
Kn\"odlseder, J., Lichti, G. G., Pl\"uschke, S., Ryan, J., Strong, A.,
\& Winkler, C. 2000, 5th Compton Symposium, AIP Conf. 510, eds. M. L.
McConnell \& J. M. Ryan (Melville, NY: AIP), 54

\bibitem[]{}
Slane, P., Hughes, J. P., Edgar, R. J., Plucinsky, P. P., Miyata, E.,
Tsunemi, H., \& Aschenbach, B. 2001a, ApJ, 548, 814

\bibitem[]{}
Slane, P., Hughes, J. P., Edgar, R. J., Plucinsky, P. P., Miyata, E.,
\& Aschenbach, B. 2001b, in Young Supernova Remnants, ed. S. S. Holt
\& U. Hwang (New York: AIP), 403

\bibitem[]{}
Spyromilio, J. 1994, MNRAS, 266, L61

\bibitem[]{}
Stathakis, R. A., Dopita, M. A., Cannon, R. D., \& Sadler, E. M.
1991, in Supernovae, ed. S. E. Woosley (New York: Springer) 95

\bibitem[]{}
Strom, R., Johnston, H. M., Verbunt, F., \& Aschenbach, B. 1995,
Nature, 373, 590

\bibitem[]{}
Truelove, J. K., \& McKee, C. F. 1999, ApJS, 120, 299

\bibitem[]{}
Tsunemi, H., Miyata, E., \& Aschenbach, B. 1999, PASJ, 51, 711

\bibitem[]{}
Tsunemi, H., Miyata, E.,  Aschenbach, B., Hiraga, J., \& Akutsu, D.
2000, PASJ, 52, 887


\bibitem[]{}
Wang, C.-Y., \& Chevalier, R. A. 2001, ApJ, 549, 1119 (WC01)



\bibitem[]{}
Winkler, P. F., Tuttle, J. H.,  Kirshner, R. P., \& Irwin, M. J.
 1988, in Supernova Remnants and the Interstellar Medium, ed. R. S.
 Roger \& T. L. Landecker (Cambridge: Cambridge Univ. Press), 65



\end{thebibliography}
\end{document}